\documentclass[runningheads]{llncs}
\usepackage{amssymb}
\usepackage{graphicx}
\usepackage{url}
\def\zombie{{\tt{sleepy}}}
\def\confused{{\tt{confused}}}
\def\motivated{{\tt{motivated}}}
\def\gentlenudge{{\it{gentle-nudge}}}
\def\hardpush{{\it{big-push}}}

\begin{document}
\title{A Conjecture Equivalent to the Collatz Conjecture}
%
%
\author{Ashish Tiwari\inst{1}\orcidID{0000-0002-5153-2686}}
%
\authorrunning{A. Tiwari}
%
\institute{Microsoft, Redmond WA 98052, USA\\
  \email{ashish.tiwari@microsoft.com}\\
  \url{http://www.csl.sri.com/users/tiwari}}
%
\maketitle              
\begin{abstract}
    We present a formulation of the Collatz conjecture that 
    is potentially more amenable to modeling and analysis 
    by automated termination checking tools.
\end{abstract}
\section{A Conjecture}
\label{sec:puzzle}

Consider an organization where we have a linear hierarchy 
among the employees.
Each employee can be in one of 3 different ``state of mind":
\zombie, \confused, or \motivated. Furthermore, 
each employee can get a \gentlenudge\ or \hardpush\ (that is
generated by their supervisor), and each employee can generate
either a \gentlenudge\ or \hardpush\ for their subordinate.

Formally, an employee is a Mealy machine~\cite{MealyWiki,Mealy}
with 3 states 
(\zombie, \confused, \motivated) and two
input (and output) symbols (\gentlenudge, \hardpush).
An organization is formally an ordered list of
employees. 
Consequently, the {\em{state}} of an organization is 
a list of states of its employees; for example,
a possible state of an organization with 3 employees
is $[\motivated, \confused, \zombie]$.

Every morning the top member of the organization gets a \gentlenudge,
which is propagated down through the day to all the employees in the hierarchy
based on the following rules:
\begin{enumerate}
    \item 
         When a \zombie\ gets a \gentlenudge, 
         they continue to remain a \zombie\ and 
         generate a \gentlenudge\ for the next person.
   \item
   When a \zombie\ gets a \hardpush, they turn \confused\ and 
   generate a \hardpush\ for the next person.
 \item
     When a \confused\ gets a \gentlenudge, they turn \zombie\ and 
   generate a \hardpush\ for the next person.
   \item
   When a \confused\ gets a \hardpush, they turn \motivated\ and 
   generate a \gentlenudge\ for the next person.
   \item
   When a \motivated\ gets a \gentlenudge, they turn \confused\ and 
   generate a \gentlenudge\ for the next person.
   \item
   When a \motivated\ gets a \hardpush, they continue to remain 
        \motivated\ and generate a \hardpush\ for the next person.
\end{enumerate}

If the last person in the hierarchy generates a \hardpush, then a new
junior \motivated\ person is hired (and added to the organization).

If at the end of the day
all members of an organization turn \zombie\
but for possibly the last member, 
then the organization goes 
{\em{bankrupt}}.

\begin{example}\label{ex1}
Let us illustrate the dynamics of an organization with an example. 
We start an organization with 3 people, who initially are in the state
    \motivated, \confused, \zombie\ in that order.
So, on Day 0, the state of the organization
    is the tuple $[ \motivated, \confused, \zombie\ ]$.

On Day 1, \motivated\ gets a \gentlenudge, so, they turn 
    \confused\ and generate a \gentlenudge, which
    turns the \confused\ into a \zombie\ and makes it
    generate a \hardpush\ for the third employee (\zombie).
    As a result, \zombie\ turns \confused\ and hires a \motivated. 
    Thus, at the end of Day 1,
    the state of the organization is 
    $
    [\confused, \zombie, \confused, \motivated].
    $

Continuing this way, we notice that at the end of Day 2, Day 3, Day 4, and Day 5
we get to states
    \begin{eqnarray*}
        {\mathtt{Day 2:}} & & [ \zombie, \confused, \motivated, \confused\ ] \\
        {\mathtt{Day 3:}} & & [ \zombie, \zombie, \motivated, \motivated\ ] \\
        {\mathtt{Day 4:}} & & [ \zombie, \zombie, \confused, \confused\ ] \\
        {\mathtt{Day 5:}} & & [ \zombie, \zombie, \zombie, \motivated\ ] 
    \end{eqnarray*}
The organization goes bankrupt on Day 5.
\end{example}

We can easily formalize the six rules enumerated above as defining the
transition relation and output function of a Mealy machine.
We write $s \rightarrow t$ to denote that
state $s$ of an organization (at the start of a day) changes to state $t$ (at the end of that day).
A state $s$ is {\em{bankrupt}} if only its last
element is possibly not \zombie.

\begin{conjecture}\label{conj1}
Every organization eventually goes bankrupt; that is, 
for any state $s$ of an organization, there is a $k \geq 0$ such that
if $s \rightarrow s_1  \rightarrow \cdots \rightarrow s_k$ is a derivation,
then $s_k$ is a backrupt state.
\end{conjecture}

\section{Conjecture~\ref{conj1} is equivalent to Collatz Conjecture}

We first recall the Collatz conjecture~\cite{CollatzWiki}.
\begin{conjecture}[Collatz Conjecture]\label{conj2}
    Let $f(n)$ be $n/2$ if $n$ is even and $3n+1$ if $n$ is odd.
    Then, for every positive natural number $n$, 
    there exists a $k\geq 0$ s.t. 
    $f^k(n) = 1$.
\end{conjecture}

We now show that Conjecture~\ref{conj1} is equivalent to the Collatz
conjecture.
We use a novel representation of numbers for this purpose. 
Each digit in our representation is either $0, 2,$ or $4$, and
the place value is interpreted as in Base-3 representation.
For example, the number $420$ in our new representation denotes
the decimal number $4*3^2 + 2*3^1 + 0*3^0$, which is $42$ in base-10.
We can only represent even numbers in this new representation.
In this new representation, if 
$0$ is written as \zombie,
$2$ as \confused, and
$4$ as \motivated, then each number $n$ can be seen as a state of
an organization. For example, $42$ is represented by the organization
$[\motivated,\confused,\zombie]$.

We modify the iterations in the statement of Collatz conjecture to go
from even number to an even number as follows: Define
$g(n)$ to be $n/2$ if $n/2$ is even, and 
$3(n/2)+1$ if $n/2$ is odd.
Now, it is an easy exercise to see that if
$m = g(n)$, and 
$s$ is a state representing $n$,
then $s \rightarrow t$ where 
$t$ represents $m$.  
This shows the equivalence to Collatz conjecture.

\begin{example}\label{ex2}
    Let us map the scenario in Example~\ref{ex1} to numbers. At the end of Day 0, we have the number $42$.
At the end of the next 5 days, we get the numbers
   $64, 32, 16, 8, 4$. Note that $4$ is represented as
   $[\zombie, \zombie, \zombie, \motivated]$. Note that
    a \zombie\ member at the head of a state
    can be dropped without affecting the numerical
    interpretation of that state.
\end{example}

There is some recent work~\cite{DBLP:journals/corr/abs-2105-14697} on trying to prove the Collatz conjecture using automated termination proving techniques for rewriting systems. For this purpose, one needs a rewriting system that mimics the iterations in Collatz conjecture. 
There is one such rewriting system in~\cite{DBLP:journals/corr/abs-2105-14697}. The formulation presented here can also be turned into a (different) rewrite system.

\begin{remark}[Open loop interpretation.]
Note that the ``control input action'' is fixed in our setting: the
    top member of the organization {\em{always}} gets a \gentlenudge\ at the start of the day.  One can 
    consider the {\em{open-loop setting}} where
    the input to the top element is a control input,
    and one could study the problem of synthesizing
a controller. We do not do so here.
\end{remark}


%
%
\bibliographystyle{splncs04}
\bibliography{main}
\end{document}